\documentclass[runningheads]{llncs}
\usepackage{graphicx}
\usepackage{amsmath}
\usepackage{mathrsfs}
\usepackage{amsfonts}
\usepackage{tikz}
\usetikzlibrary{automata,positioning}
\usepackage{color}
\usepackage{dirtytalk}
\usepackage{algorithm,algorithmicx,algpseudocode}
\usepackage{graphicx}
\usepackage{dirtytalk}
\usepackage{amsmath,amssymb,amsfonts}
\usepackage{enumitem}
\usepackage{graphicx}

\tikzset{every state/.style={minimum size=0.5pt}}
\tikzset{every edge/.append style={font=\small}}

\newcommand{\ATAC}{\texttt{ATAC} }
\newcommand{\automaton}{\mathscr{A}}
\newcommand{\automatonset}{\automaton  = (L, l_0, \Sigma, C, I, T)}

\algnewcommand\Or{\textbf{or}}
\algnewcommand\algorithmicforeach{\textbf{for each}}
\algdef{S}[FOR]{ForEach}[1]{\algorithmicforeach\ #1\ \algorithmicdo}

\begin{document}
\title{\ATAC: A Tool for Automating Timed~Automata~Construction\thanks{This work has received funding from the European Union's Horizon 2020 research and innovation programme under the Marie Sk\l{}odowska-Curie grant agreement No 798482.}}
\author{Beyazit Yalcinkaya
%\inst{1}
%\orcidID{0000-0001-9987-635X}
\and
Ebru Aydin Gol
%\inst{1}
%\orcidID{0000-0002-5813-9836}
}
\authorrunning{B. Yalcinkaya and E. Aydin Gol}
\institute{Middle East Technical University, Ankara, Turkey
\email{\{beyazit.yalcinkaya,ebrugol\}@metu.edu.tr}}
\maketitle              % typeset the header of the contribution
\begin{abstract}
In this paper, we focus on the design and verification of timed automata (TA).
We introduce a new method for assisting construction and verification of TA models along with a tool implementing the proposed method, i.e., \ATAC: \emph{Automated Timed Automata Construction}.
Our method provides two main functionalities, i.e., construction of TA models from descriptions and generation of temporal logic queries from specifications.
Both description and specification sentences shall follow our well-defined structured natural language definition.
TA models constructed from descriptions and temporal logic queries generated from specifications can be imported to UPPAAL, a verification tool for TA models.
The goal is to accelerate the design phase for real-time systems by assisting the construction and verification of a formal model. We believe \ATAC can be useful especially during the initial phases of the design process and help designers to avoid erroneous models.
\end{abstract}

\section{Introduction}

\emph{Cyber-physical systems} (CPS) are everywhere and they are getting increasingly more attention from the research community as well as the industry with the advancements in the digital transformation and increase in the adaptation of intelligent systems. 
As their application areas increase, models representing CPS are getting more complex and sophisticated.
Systems with timing requirements, i.e., \emph{real-time systems} (RTS), are an essential part of CPS. 
\emph{Timed automata} (TA) \cite{Alur:CPSBook,alur1993model,alur1994theory}, a widely-used and well-known RTS formalism, is used for modeling, designing, and verifying CPS. 
From railroad crossing systems \cite{Heitmeyer:94,Wang:2004} to cardiac pacemakers \cite{pacemakers:2015} to scheduling of real-time systems \cite{david2009model,guan2007,yalcinkaya2019exact}, TA are used to model and analyze numerous systems.
Algorithms for verifying TA against formal specifications such as temporal logics exist and implemented in various tools including UPPAAL \cite{UPPAAL4:0}, a tool for modeling, designing, and verifying TA models.

As CPS get more complicated, designing reliable TA models is getting tedious and error-prone. 
Since most of the RTS are safety-critical, errors in the design of TA models are not tolerable.
%Moreover, designing models that can be applied to reasonably-sized systems requires even more expertise.
%In a recent paper \cite{yalcinkaya2019exact}, it has been shown that some of the TA based schedulability tests for real-time tasks are not applicable to even small-sized task sets due to state-space explosion problem caused by complicated design of models.
However, since there is no automated way of TA design and no well-defined design procedure to follow, designing TA models inherently requires human creativity and knowledge, which is an unavoidable source of error and considerably dangerous for safety-critical systems.

\paragraph*{This paper}
We introduce novel methods to assist TA design. % The developed methods are implemented as a tool named \emph{\ATAC (Automated Timed Automata Construction)}.
% a new tool, \emph{\ATAC (Automated Timed Automata Construction)}, to assist TA design.
We generate TA models and temporal logic formulas from specifications and system descriptions given in structured natural language in an automated way. 
We restrict description and specification sentences to a limited set of English sentences to simplify the design descriptions.
Given a description sentence, any implied timing, synchronization, and transition information about the system are extracted, then a TA model following these descriptions is constructed. For the specification sentences, specifications are mapped to formulas in the UPPAAL's query language, a subset of Timed Computation Tree Logic (TCTL). If there is a specific timing behavior implied by the given sentence, the necessary clocks are placed on the model so that the corresponding formula can be verified.  That is, a non-trivial mapping is performed from input sentences to TA models and temporal logic queries by taking care of low-level design choices such as clock allocations, synchronization of TA models, and generation of temporal logic formulas.
The goal is to assist development of TA models, not to fully automate the whole design process. Although, the developed methods allow users to define complete TA models, we foresee that for a complex model, description sentences would be too long and it would not be  simpler than manually designing the model. For this reason, we believe the proposed methods would be useful to construct a skeleton for the complex model together with the placement of necessary clocks for the verification of timing properties, and then the user can take over to add the task-specific components of the model. The developed methods are implemented as a Python program named \emph{\ATAC (Automated Timed Automata Construction)}. Given the system descriptions and specifications,  \ATAC outputs an XML file and a query file which can be imported to UPPAAL.% for verification.

\paragraph*{Related Work}
To the best of out knowledge, the proposed method is the first attempt to automate TA construction at this level.
Although, there has been some work on different aspects of TA generation such as parameter synthesis~\cite{inversemethod} and controller synthesis~\cite{TAcontrol}, no prior work has been focused on the automation of TA design.
In \cite{andersen1995automatic}, the authors present a theoretic approach to the problem of automatic synthesis of real-time systems. In their solution, from a given set of timed models and a real-time logical specification, they check the existence of a real-time model which can yield a TA network satisfying given specification.
In LTLMop Project \cite{ltlmop}, a set of sentences are mapped to Linear Temporal Logic (LTL) specifications for robot control by defining a formal grammar.
In \cite{andre2012imitator}, \textsf{IMITATOR} is presented which is a tool for parametric verification and robustness analysis of parametric timed automata.
In \cite{enoiu2013model}, a transformation method from programmable logic controllers represented with functional block diagrams to a TA model has been introduced.
However, no work has been done on assisting TA design at the level of structured natural language.
%Although, their problem was in a different domain, usage of a formal grammar for translation of English sentences to technical concepts and implementing a tool for the problem contain a similar notion to some aspects of our approach.
%However, our sentences are at a higher-level, i.e., they hide most of the implementation details.

The paper is outlined as follows. Sec.~\ref{prelim} presents basic definitions of TA and UPPAAL.
Sec.~\ref{desc} and Sec.~\ref{spec} present details of the TA construction and specification generation, respectively.
Sec.~\ref{tool} reveals details of \ATAC. Sec.~\ref{case} elaborates a case study, and Sec.~\ref{conclusion} concludes the paper.

\section{Timed Automata}\label{prelim}

A timed automaton (TA)~\cite{Alur:CPSBook,alur1994theory,larsen1993time} is a finite-state machine extended with a finite set of real-valued clocks progressing monotonically and measuring the time spent after their latest resets. $\Phi(C)$ is a set of clock constraints over a set of clocks $C$. A \textit{clock constraint} $\phi \in \Phi(C)$ is given by the grammar: $\phi ::= x  \leq c \mid x \geq c \mid x < c \mid x > c \mid \phi \land \phi$, where $x \in C$ and $c \in \mathbb{Q}$.

\begin{definition}[Timed Automata]
A \textit{timed automaton}  $\automatonset$ is a tuple, where (i) $L$ is a finite set of locations, (ii) $l_{0} \in L$ is an initial location, (iii) $\Sigma$ is a finite set of labels, (iv) $C$ is a finite set of clocks, (v) $I$ is a mapping from $L$ to $\Phi(C)$, and (vi) $T \subseteq L \times \Sigma \times 2^{C} \times \Phi(C) \times L$ is a set of transitions. $t = (l_s, \alpha, \lambda, \phi, l_t) \in T$ is a transition from $l_s$ to $l_t$ on symbol $\alpha$. $\lambda$ is a set of clocks reset to zero on $t$ and $\phi$ is a clock constraint tested for enabling $t$.
\end{definition}

UPPAAL \cite{UPPAAL4:0,larsen1997uppaal}, is a tool for modeling, designing, simulating, and verifying TA. UPPAAL extends the TA formalism % with urgent and committed locations, synchronization channels, programming structures etc. 
to provide wider implementation opportunities for RTS. One such extension is the synchronization channels that can be added to transitions to synchronize two TA, i.e., the synchronized TAs have to take the transition at the same time. The developed methods support the synchronization channels as it is fundamental to model network of RTS. However, other extensions are omitted to avoid complicated descriptions.

\section{Timed Automata Construction}\label{desc}

% One of the two main aspects of our method is TA construction.
In this section, we define proposed approach to generate a TA model from high level descriptions of a RTS.  First, we define a formal grammar for the system description sentences. For a set of sentences following this grammar, we construct a TA by allocating locations, transitions, clocks, and synchronization channels by inferring details of the model in an automated way. 
% If a given set of sentences follows this grammar, then we generate a TA model for the system in an automated way. 
% The tool implementing the proposed approach, \ATAC, generates an XML file for the resulting TA that can be imported to UPPAAL. 
% Given a set of high-level description sentences in structured natural language, we construct specified TA model by allocating locations, transitions, clocks, and synchronization channels by inferring details of the model.
% then our tool constructs the implied TA model and outputs an XML file that can be imported to UPPAAL.
Eqn.~\eqref{desc_gram} defines main grammar rules for the TA construction.
Notice that, in this definition, \textbf{A} indicates name of a TA model, \textbf{L} indicates name of a location, \textbf{S} indicates name of a synchronization channel, italic font indicates strings appearing in the rules, and $\epsilon$ is the empty string. A set of sentences for describing a TA model should follow the rule $\phi_{TA}$.

%\EqnSpace
%\EqnSpace

To define a TA (\ref{ta}), first, an \emph{initialization sentence} following $\phi_{init}$ should be given. It declares name of the model and its locations.
If the model contains a single location (\ref{init1}), otherwise (\ref{init2}) should be used.
\emph{System sentences} (\ref{sys}) follow the initialization sentence.
These sentences can refer to either invariance properties (\ref{inv}) or transitions (\ref{t1}-\ref{t6}) of the model.
%An invariant is a formula that is ensured to be true for a given location.

\begin{subequations}\label{eq:taall}
\begin{align}
	\phi_{TA} \  ::= &\ \phi_{init} \ \phi_{sys}\label{ta}\\
	\phi_{init} \  ::= &\ \textbf{A } \textit{can only be } \textbf{L}\label{init1}\\
			  \mid &\ \textbf{A } \textit{can be } \phi_{locs} \ \textit{and it is initially } \textbf{L}\label{init2}\\
	\phi_{sys} \ ::= &\ \phi_{invrt} \ \phi_{sys} \ \mid \ \phi_{tran} \ \phi_{sys}\ \mid \epsilon\label{sys}\\
	\phi_{invrt} \ ::= &\ \textit{for } \textbf{A } \phi_{ic} \ \textit{in } \textbf{L} \ \mid \ \textit{for } \textbf{A } \textit{the time spent in } \textbf{L } \textit{cannot be } \phi_{icons}\label{inv}\\
	\phi_{tran} \ ::= &\ \textbf{A } \textit{can } \phi_{go}\label{t1}\\
			\mid &\ \textbf{A } \textit{can send } \textbf{S } \textit{and } \phi_{go}\label{t2}\\
			\mid &\ \textit{if } \phi_{sc} \ \textit{then } \textbf{A } \textit{can } \phi_{go}\label{t3}\\
			\mid &\ \textit{if } \phi_{tc} \ \textit{then } \textbf{A } \textit{can } \phi_{go}\label{t4}\\
			\mid &\ \textit{if } \phi_{tc} \ \textit{then } \textbf{A } \textit{can send } \textbf{S } \textit{and } \phi_{go}\label{t5}\\
			  \mid &\ \textit{if } \phi_{sc} \ \textit{and } \phi_{tc} \ \textit{then } \textbf{A } \textit{can } \phi_{go}\label{t6}
\end{align}
\label{desc_gram}
\end{subequations}

Grammar defines six distinct transition types:
\emph{Simple transition} (\ref{t1}) defines a transition from a set of locations to a set of locations.
\emph{Synchronization transition} (\ref{t2}) defines a simple transition sending a synchronization signal.
\emph{Synchronization conditional simple transition} (\ref{t3}) defines a simple transition which can only be taken if a synchronization signal is received.
\emph{Time conditional simple transition} (\ref{t4}) defines a simple transition which can only be taken if a condition on clocks is satisfied.
\emph{Time conditional synchronization transition} (\ref{t5}) defines a synchronization transition which can only be taken if a condition on clocks is satisfied.
\emph{Synchronization-time conditional simple transition} (\ref{t6}) defines a simple transition which can only be taken if a synchronization signal is received and a condition on clocks is satisfied.
% A trivial is that if there exists a seventh transition type named \emph{synchronization-time conditional synchronization transition}. Such a transition type is not supported by UPPAAL that is why it is not included in the grammar.

Eqn.~\eqref{help_gram} presents helper grammar rules.
\emph{Locations} (\ref{locs}) defines a set of locations in the model; \emph{synchronization condition} (\ref{sc}) indicates a synchronization condition; \emph{transition condition} (\ref{tc}) describes a time condition implicitly referring existence of some clocks without giving any low-level details; \emph{invariant condition} (\ref{ic}) is similar to a transition condition as it also defines a time condition but for a location invariant; \emph{time constraint} (\ref{tcons}) describes a timing constraint by relational operators; \emph{invariant constraint} (\ref{icons}) defines a timing constraint by using only more than and more than or equal to relational operators; \emph{equality} (\ref{eq}) indicates either existence or nonexistence of the equality case for relational operators; \emph{entering or leaving} (\ref{el}) indicates if the transition under consideration is either entering or leaving mentioned location in the sentence; and finally, \emph{go} (\ref{go}) indicates transitions from a set of locations to a set locations.

%\EqnSpace
%\EqnSpace

Note that clocks are not explicitly defined in the grammar. For \emph{transition condition}s and \emph{invariant condition}s, the clock allocation, reset placement, and constraint generation are performed in an automated way. In particular, for each transition and invariant condition appearing in the description, a new clock is created. If the sentence refers to a condition depending on entering a location, then this new clock is reset on each transition that ends in the given location. If the sentence refers to a condition depending on leaving a location, then the clock is reset on each transition that leaves the given location. 
This entering/leaving information for resetting clocks is inferred from $\phi_{el}$ rule in the grammar. To generate conditions of a clock, we extract the information given by $\phi_{tc}$ and $\phi_{ic}$. 
% The necessary clocks are generated and the  , i.e., \ATAC infers existence of a clock and allocates a clock for the given timing constraint

\begin{subequations}
\begin{align}
	\phi_{locs} \  ::= &\ \textbf{L } \mid \ \textbf{L } \phi_{locs} \label{locs}\\
	\phi_{sc} \ ::= &\ \textbf{S } \textit{is received} \label{sc}\\
	\phi_{tc} \ ::= &\ \textit{the time spent after } \phi_{el} \ \textbf{L } \textit{is } \phi_{tcons} \ \mid \ \phi_{tc} \ \textit{and } \phi_{tc} \label{tc}\\
	\phi_{ic} \ ::= &\ \textit{the time spent after } \phi_{el} \ \textbf{L } \textit{cannot be } \phi_{icons} \ \mid \ \phi_{ic} \ \textit{and } \phi_{ic} \label{ic}\\
	\phi_{tcons} \ ::= &\ \textit{more than } \phi_{eq} \ \textbf{N} \ \mid \ \textit{less than } \phi_{eq} \ \textbf{N }\nonumber \\
	 \mid & \ \textit{equal to } \ \textbf{N }
	\mid \ \phi_{tcons} \ \textit{and } \phi_{tcons} \label{tcons}\\
	\phi_{icons} \ ::= &\ \textit{more than } \phi_{eq} \ \textbf{N } \mid \ \phi_{icons} \ \textit{and } \phi_{icons} \label{icons}\\
	\phi_{eq} \ ::= &\ \textit{or equal to} \ \mid \ \epsilon \label{eq}\\
	\phi_{el} \ ::= &\ \textit{entering } \ \mid \ \textit{leaving } \label{el}\\
	\phi_{go} \ ::= &\ \textit{go from } \phi_{locs} \ \textit{to } \phi_{locs}\label{go}
\end{align}
\label{help_gram}
\end{subequations}

It is ensured that ordering of the description sentences is not significant, i.e., a given set of sentences following the description grammar is mapped to a unique TA  for any ordering. Once all the description sentences are analyzed for a TA, the clock reduction algorithm from~\cite{yalcinkaya2019clock} is applied to reduce the number of clocks. As each timing constraint is, initially, encoded with a separate clock, the clock reduction step significantly reduces the number of clocks, which is essential to reduce the verification time and improve maintainability of the model. 
% Using the rules of the description grammar~\eqref{desc_gram} , a TA model can be constructed. 
% A detailed example representing the usage of these rules is given in the Sec.~\ref{case}. 

\section{Specification Generation}\label{spec}

In this section, we define the structure for the specification sentences, the mapping to the TCTL formulas and the corresponding query in the UPPAAL's query language.
% Another main functionality of \ATAC is specification generation.
% Given a specification sentence in structured natural language, \ATAC can map the sentence to a TCTL formula within the scope of UPPAAL's query language.
In addition to formula generation, 
% \ATAC performs more than a mapping between sentences to TCTL formulas, i.e., 
for a specification with a timing condition, we place a new clock on the model and generate the formula with respect to this new clock.
That is, the user does not specify anything regarding a new clock and its value (similar to the description grammar). Only a high-level specification is given by the user and the necessary clock placement is performed automatically for the verification of the inferred formula.
We define the structured natural language sentences for generating queries with a formal grammar which entirely covers UPPAAL's query language.
% Below, we present the formal grammar for the specification generation.
Notice that, the same convention as in the specification grammar and some of the helper rules are used.
A specification sentence should follow the rule $\phi_{spec}$ in Eqn.~\ref{spec_gram}.
% \EqnSpace
%\EqnSpace

Four distinct specification types are defined by the grammar:
\emph{General} (\ref{general}) defines the general format of UPPAAL's query language which is a subset of TCTL. Since UPPAAL does not allow nested path formulas, the general query format can be realized by a path formula ($\phi_{pf}$) followed by a state formula ($\phi_{sf}$), which is further explained below.
\emph{Deadlock} (\ref{deadlock}) indicates the special case of UPPAAL's query language for checking the existence of a deadlock. This rule is mapped to the following TCTL formula: \say{$A\square \neg deadlock$} which is written as \say{\texttt{A[] not deadlock}} in UPPAAL.
\emph{Leads to} (\ref{leads_to}) is again a special case of UPPAAL's query language. A sentence following this rule basically says that a state formula leads to another state formula. Suppose a sentence following (\ref{leads_to}) is given as ``$\psi_1$ leads to $\psi_2$'', then this sentence is mapped to the following TCTL formula: $A\square (\psi_1 \implies A\diamond \psi_2)$ which is written as \say{\texttt{$\psi_1$ --> $\psi_2$}} in UPPAAL.
\emph{Shorthand} (\ref{s1}) is a shorthand rule that we introduced for convenience. It indicates that a given location shall always be visited within a specific amount of time. Suppose a sentence is given following this rule and it belongs to a TA named Template, name of the inferred location is Location, and the amount of time specified is $4$, then the tool maps it to the following TCTL formula: \say{$A\square \ \neg Template.Location \ \lor \ x \leq 4$} which is written as \say{\texttt{A[] not Template.Location or x <= 4}} in UPPAAL. Notice that \texttt{x} is a clock variable that is automatically created and it is reset to zero in every transition leaving given location.

\begin{subequations}\label{eq:spec_grammar}
\begin{align}
	\phi_{spec} \  ::= &\ \textit{it } \phi_{pf} \ \textit{be the case that }\phi_{sf} \label{general}\\
	\mid &\ \textit{deadlock never occurs} \label{deadlock}\\
	\mid &\ \phi_{sf} \ \textit{leads to } \phi_{sf} \label{leads_to}\\
	\mid &\ \textit{for } \textbf{A } \textbf{L } \textit{shall hold within every } \textbf{N} \label{s1}\\
	\phi_{pf} \ ::= &\ \textit{shall always } \mid \ \textit{shall eventually}\nonumber \\
	 \mid & \ \textit{might always } \mid \ \textit{might eventually}  \label{pf}\\
	\phi_{sf} \ ::= &\ \phi_{atom} \ \mid \ \phi_{atom} \ \phi_{op} \ \phi_{sf}  \label{sf}\\
	\phi_{atom} \ ::= &\ \textit{for } \textbf{A } \textit{the time spent after } \phi_{el} \ \textbf{L } \textit{is } \phi_{tcons} \nonumber\\
	\mid &\ \textit{for } \textbf{A } \phi_{locs} \ \textit{holds} \ \mid \ \textit{for } \textbf{A } \phi_{locs} \ \textit{does not hold}\label{atom}\\
	\phi_{op} \ ::= &\ \textit{and} \ \mid \ \textit{or} \ \mid \ \textit{implies}\nonumber
\end{align}
\label{spec_gram}
\end{subequations}

\emph{Path formulas} (\ref{pf}) are mapped to TCTL path formulas as follows: \textit{shall always} is mapped to $A\square$, \textit{shall eventually} is mapped to $A\diamond$, \textit{might always} is mapped to $E\square$, and \textit{shall eventually} is mapped to $E\diamond$. \emph{State formulas} (\ref{sf})  can either be an atomic proposition or some atomic propositions connected by conjunction, disjunction, and implication. \emph{Atomic propositions} (\ref{atom}) can infer to either a time condition or visiting a location of the model.

For each time condition, a unique clock is generated. The clock reduction process is not applied to these clocks. They are only reset to zero on necessary transitions of the model and only checked in the generated queries. This functionality allows users to avoid manual clock allocations for the verification of specifications. %, i.e., to check a model against a specification, user does not need to modify model. 

\section{Implementation of \ATAC}\label{tool}

TA and formula construction methods presented in this paper are implemented as a Python program called \ATAC~\cite{ATAC}.  The user can give multiple TA descriptions ($\phi_{TA}$~\eqref{eq:taall}) to construct a network of TA. Descriptions of these TA can intersect, i.e., TA descriptions does not need to be sequential. \ATAC handles allocation of locations, transitions, clocks, and synchronization channels automatically as described in Sec.~\ref{desc} and Sec.~\ref{spec}. The tool also implements the clock reduction algorithm from~\cite{yalcinkaya2019clock} that reduces the number of clocks via clock renaming. As described in Sec.~\ref{desc}, initially a new clock is generated for each sentence with a time condition, which results in a large number of clocks. Then, the reduction is performed to reduce the number of clocks. After TA construction, clock reduction, and specification mapping, the resulting TA is mapped to an XML file and its specifications are mapped to a query file. Both of these files can be imported to UPPAAL. % Pyuppaal \cite{PyUppaal} module is used for mapping TA models to XML.

\section{A Case Study: Train-Gate TA}\label{case}

\begin{figure*}[t]
\begin{center}
\includegraphics[width=0.70\linewidth]{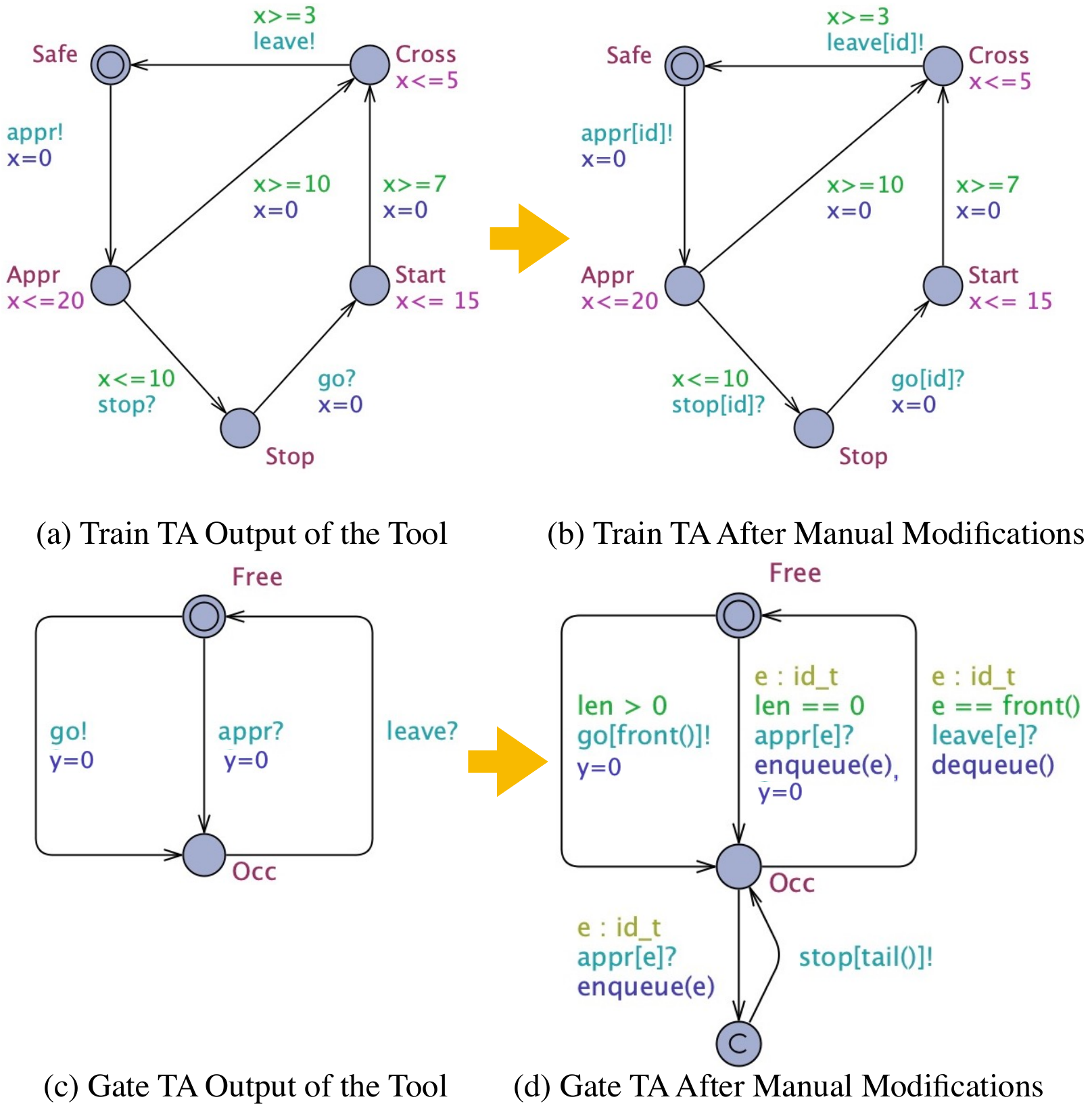}
\end{center}
\vspace{-0.5cm}
\caption{Train-Gate Model}\label{ex2}
\end{figure*}

In this section, we present a case study to illustrate the proposed methods and the intended usage of \ATAC. We picked a well-know example which is also provided in the installation package of UPPAAL as a demo example, i.e., Train-Gate model (Fig.~\ref{ex2}). This model consists of two TA models: a train and a gate. A certain number of the train model can be instantiated and each instance of the train TA is given an identification number ($id$). There is a single instantiation for the gate TA controlling which train to pass the gate and which ones to wait. The communication between the gate and trains is performed by an array of synchronization channels. The channel array is indexed by train ids. In addition to the channel array, the model uses other extensions implemented in UPPAAL such as nondeterministic selection of integers, C-like programming structures, and committed locations.

\subsection{Description of Train TA}
Below, we list description sentences to construct the model given in Fig.~\ref{ex2}-(a). % As it can be seen, only ten sentences are enough to construct such a Train TA. 
Notice that with the clock reduction algorithm, our tool automatically reduces the number of clocks to one for multiple timing requirements on the model. Fig.~\ref{ex2}-(b) presents the TA model obtained after manual modification of the automatically constructed model. To obtain the final model, only some C-like programming structures and channel indexes are added to the model manually.
\begin{itemize}[nolistsep]
	\item Train can be Safe Appr Cross Stop Start and it is initially Safe.
	\item Train can send Appr and go from Safe to Appr.
	\item If the time spent after entering Appr is more than or equal to 10, then Train can go from Appr to Cross.
	\item If Stop is received and the time spent after entering Appr is less than or equal to 10, then Train can go from Appr to Stop.
	\item If Go is received, then Train can go from Stop to Start.
	\item If the time spent after entering Start is more than or equal to 7, then Train can go from Start to Cross.
	\item If the time spent after entering Cross is more than or equal to 3, then Train can send Leave and go from Cross to Safe.
	\item For Train, the time spent in Appr cannot be more than 20.
	\item For Train, the time spent in Start cannot be more than 15.
	\item For Train, the time spent in Cross cannot be more than 5.
\end{itemize}

\subsection{Description of Gate TA}
Below, we list description sentences for Gate TA given in Fig.~\ref{ex2}-(c). Four sentences are needed for this model. Fig.~\ref{ex2}-(d) presents the TA model obtained after manual modification of the automatically constructed model. To obtain the final model, only nondeterministic selection for integer variables and a committed location are added to the model manually.
\begin{itemize}[nolistsep]
	\item Gate can be Free Occ and it is initially Free.
	\item Gate can send Go and go from Free to Occ.
	\item If Appr is received, then Gate can go from Free to Occ.
	\item If Leave is received, then Gate can go from Occ to Free.
\end{itemize}

\subsection{Specifications for the whole model}
Below, we give some useful specifications and their resulting UPPAAL queries.
\begin{itemize}[nolistsep]
	\item It might eventually be the case that for Gate, Occ holds:\texttt{E<> Gate.Occ}
	\item For Gate, Free holds leads to for Train, Cross holds: \texttt{Gate.Free --> Train.Cross}
	\item It shall always be the case that for Train, Cross does not hold or for Gate, Free does not hold: \texttt{A[] not Train.Cross or not Gate.Free}
	\item Deadlock never occurs: \texttt{A[] not deadlock}
	\item For Gate, Free shall hold within every 40: \texttt{A[] not Gate.Free or y <= 40}
\end{itemize}

Notice that, in Gate TA, clock \texttt{y} does not exist in the original example of UPPAAL. Since the last specification has a timing condition, it is generated automatically to verify the specification.

\section{Conclusion}\label{conclusion}

In this paper, we presented a new method for assisting TA construction and a tool implementing this method. \ATAC constructs TA models and generates queries from high-level descriptions and specifications given in structured natural language. Output models and formulas can be imported to UPPAAL for verification. We gave a brief introduction to the basic concepts of TA and UPPAAL, and then presented description and specification grammars. Finally, we presented a case study generated with \ATAC. Additional examples are included in the tool~\cite{ATAC}. % elaborated case studies. %  of \ATAC and revealed implementation details.
Although our method can be used for construction of a complete TA along with queries for specifications, we suggest
% suggested usage of \ATAC is 
using it as an assistance tool rather than a fully automated one. We encourage using it in the initial phases of the design where only a textual description of the model under construction is available. This way our method can provide an almost complete model which can be further improved. In addition, it identifies the necessary number of clocks and allocates them on the TA model for the considered RTS. Furthermore, accompanying the model with the specifications and descriptions given in natural language improves its maintainability.

\bibliographystyle{IEEEtran}
\bibliography{ebru_t}

\begin{thebibliography}{10}
\providecommand{\url}[1]{#1}
\csname url@samestyle\endcsname
\providecommand{\newblock}{\relax}
\providecommand{\bibinfo}[2]{#2}
\providecommand{\BIBentrySTDinterwordspacing}{\spaceskip=0pt\relax}
\providecommand{\BIBentryALTinterwordstretchfactor}{4}
\providecommand{\BIBentryALTinterwordspacing}{\spaceskip=\fontdimen2\font plus
\BIBentryALTinterwordstretchfactor\fontdimen3\font minus
  \fontdimen4\font\relax}
\providecommand{\BIBforeignlanguage}[2]{{%
\expandafter\ifx\csname l@#1\endcsname\relax
\typeout{** WARNING: IEEEtran.bst: No hyphenation pattern has been}%
\typeout{** loaded for the language `#1'. Using the pattern for}%
\typeout{** the default language instead.}%
\else
\language=\csname l@#1\endcsname
\fi
#2}}
\providecommand{\BIBdecl}{\relax}
\BIBdecl

\bibitem{Alur:CPSBook}
R.~Alur, \emph{Principles of Cyber-Physical Systems}.\hskip 1em plus 0.5em
  minus 0.4em\relax The MIT Press, 2015.

\bibitem{alur1993model}
R.~Alur, C.~Courcoubetis, and D.~Dill, ``Model-checking in dense real-time,''
  \emph{Information and computation}, vol. 104, no.~1, pp. 2--34, 1993.

\bibitem{alur1994theory}
R.~Alur and D.~L. Dill, ``A theory of timed automata,'' \emph{Theoretical
  computer science}, vol. 126, no.~2, pp. 183--235, 1994.

\bibitem{Heitmeyer:94}
C.~Heitmeyer and N.~Lynch, ``The generalized railroad crossing: a case study in
  formal verification of real-time systems,'' in \emph{1994 Proceedings
  Real-Time Systems Symposium}, Dec 1994, pp. 120--131.

\bibitem{Wang:2004}
F.~Wang, ``Formal verification of timed systems: a survey and perspective,''
  \emph{Proceedings of the IEEE}, vol.~92, no.~8, pp. 1283--1305, Aug 2004.

\bibitem{pacemakers:2015}
M.~Kwiatkowska, A.~Mereacre, N.~Paoletti, and A.~Patan{\`e}, ``Synthesising
  robust and optimal parameters for cardiac pacemakers using symbolic and
  evolutionary computation techniques,'' in \emph{Hybrid Systems Biology},
  A.~Abate and D.~{\v{S}}afr{\'a}nek, Eds.\hskip 1em plus 0.5em minus
  0.4em\relax Cham: Springer International Publishing, 2015, pp. 119--140.

\bibitem{david2009model}
A.~David, J.~Illum, K.~G. Larsen, and A.~Skou, ``Model-based framework for
  schedulability analysis using {UPPAAL} 4.1,'' in \emph{Model-based design for
  embedded systems}, 2009, pp. 117--144.

\bibitem{guan2007}
N.~Guan, Z.~Gu, Q.~Deng, S.~Gao, and G.~Yu, ``Exact schedulability analysis for
  static-priority global multiprocessor scheduling using model-checking,'' in
  \emph{Proc. of SEUS}, 2007, pp. 263--272.

\bibitem{yalcinkaya2019exact}
B.~Yalcinkaya, M.~Nasri, and B.~B. Brandenburg, ``An exact schedulability test
  for non-preemptive self-suspending real-time tasks,'' \emph{IEEE/ACM Design,
  Automation and Test in Europe (DATE)}, 2019.

\bibitem{UPPAAL4:0}
G.~Behrmann, A.~David, K.~G. Larsen, J.~Hakansson, P.~Petterson, W.~Yi, and
  M.~Hendriks, ``Uppaal 4.0,'' in \emph{Proceedings of the 3rd International
  Conference on the Quantitative Evaluation of Systems}, ser. QEST '06.\hskip
  1em plus 0.5em minus 0.4em\relax Washington, DC, USA: IEEE Computer Society,
  2006, pp. 125--126.

\bibitem{inversemethod}
E.~Andr\'e, T.~Chatain, L.~Fribourg, and E.~Encrenaz, ``An inverse method for
  parametric timed automata,'' \emph{International Journal of Foundations of
  Computer Science}, vol.~20, no.~05, pp. 819--836, 2009.

\bibitem{TAcontrol}
E.~Asarin and O.~Maler, ``As soon as possible: Time optimal control for timed
  automata,'' in \emph{Hybrid Systems: Computation and Control}, F.~W.
  Vaandrager and J.~H. van Schuppen, Eds.\hskip 1em plus 0.5em minus
  0.4em\relax Berlin, Heidelberg: Springer Berlin Heidelberg, 1999, pp. 19--30.

\bibitem{andersen1995automatic}
J.~H. Andersen, K.~J. Kristoffersen, K.~G. Larsen, and J.~Niedermann,
  ``Automatic synthesis of real time systems,'' in \emph{International
  Colloquium on Automata, Languages, and Programming}.\hskip 1em plus 0.5em
  minus 0.4em\relax Springer, 1995, pp. 535--546.

\bibitem{ltlmop}
\BIBentryALTinterwordspacing
\relax Cornell Verifiable Robotics Research~Group, ``Ltlmop project page,''
  2013. [Online]. Available: \url{https://ltlmop.github.io/}
\BIBentrySTDinterwordspacing

\bibitem{andre2012imitator}
{\'E}.~Andr{\'e}, L.~Fribourg, U.~K{\"u}hne, and R.~Soulat, ``Imitator 2.5: A
  tool for analyzing robustness in scheduling problems,'' in
  \emph{International Symposium on Formal Methods}.\hskip 1em plus 0.5em minus
  0.4em\relax Springer, 2012, pp. 33--36.

\bibitem{enoiu2013model}
E.~P. Enoiu, D.~Sundmark, and P.~Pettersson, ``Model-based test suite
  generation for function block diagrams using the uppaal model checker,'' in
  \emph{2013 IEEE Sixth International Conference on Software Testing,
  Verification and Validation Workshops}.\hskip 1em plus 0.5em minus
  0.4em\relax IEEE, 2013, pp. 158--167.

\bibitem{larsen1993time}
K.~G. Larsen and W.~Yi, ``Time abstracted bisimulation: Implicit specifications
  and decidability,'' in \emph{International Conference on Mathematical
  Foundations of Programming Semantics}.\hskip 1em plus 0.5em minus 0.4em\relax
  Springer, 1993, pp. 160--176.

\bibitem{larsen1997uppaal}
K.~G. Larsen, P.~Pettersson, and W.~Yi, ``Uppaal in a nutshell,''
  \emph{International Journal on Software Tools for Technology Transfer
  (STTT)}, vol.~1, no.~1, pp. 134--152, 1997.

\bibitem{yalcinkaya2019clock}
B.~Yalcinkaya and E.~A. Gol, ``Clock reduction in timed automata while
  preserving design parameters,'' in \emph{2019 IEEE/ACM 7th International
  Conference on Formal Methods in Software Engineering (FormaliSE)}.\hskip 1em
  plus 0.5em minus 0.4em\relax IEEE, 2019, pp. 31--40.

\bibitem{ATAC}
B.~Yalcinkaya, ``Atac,'' \url{https://github.com/beyazit-yalcinkaya/atac},
  2020.

\end{thebibliography}

\end{document}